# Two-scale momentum theory for time-dependent modelling of large wind farms

**Takafumi Nishino**[1,*] **and Thomas D. Dunstan**[2]

[1]Department of Engineering Science, University of Oxford, Parks Road, Oxford OX1 3PJ, UK

[2]Met Office, FitzRoy Road, Exeter EX1 3PB, UK

This paper presents a theory based on the law of momentum conservation to define and help analyse the problem of large wind farm aerodynamics. The theory splits the problem into two sub-problems; namely an 'external' (or farm-scale) problem, which is a time-dependent problem considering large-scale motions of the atmospheric boundary layer (ABL) to assess the amount of momentum available to the ABL's bottom resistance at a certain time; and an 'internal' (or turbine-scale) problem, which is a quasi-steady (in terms of large-scale motions of the ABL) problem describing the breakdown of the ABL's bottom resistance into wind turbine drag and land/sea surface friction. The two sub-problems are coupled to each other through a non-dimensional parameter called 'farm wind-speed reduction factor' or 'farm induction factor,' for which a simple analytic equation is derived that can be solved iteratively using information obtained from both sub-problems. This general form of coupling allows us to use the present theory with various types of flow models for each scale, such as a numerical weather prediction (NWP) model for the external problem and a computational fluid dynamics (CFD) model for the internal problem. The theory is presented for a simplified wind farm situation first, followed by a discussion on how the theory can be applied (in an approximate manner) to real-world problems; for example, how to estimate the power loss due to the so-called 'wind farm blockage effect' for a given large wind farm under given environmental conditions.

**Key words:**

## 1. Introduction

The aerodynamic performance of a large array of wind turbines, or a wind farm, depends on both natural and technological factors at various scales, ranging from regional weather conditions, through the layout of turbine array, down to detailed rotor design and operating conditions of each individual turbine. Because of this multi-scale nature, the problem of wind farm aerodynamics is usually split into a few sub-problems, such as regional-scale, array-scale and turbine-scale problems, to investigate key flow physics at each scale. The challenge here is to consider the effect of inter-scale interactions appropriately, which is crucial for future 'high-level' optimisation of large wind farms examining not only the layout but also the design and operating conditions of turbines simultaneously (Nishino and Hunter 2018). In this paper we propose a simple theory based on the law of momentum conservation that allows us to split the problem of wind farm aerodynamics into external (farm-scale) and internal (turbine-scale) sub-problems and to describe their relationship in a generic manner, i.e., regardless of the specific details of flow models employed at each scale.

One of the motivations behind the present theoretical work is to provide a basis for estimating the loss of wind farm power due to the so-called wind farm blockage effect (Bleeg et al. 2018), i.e., the effect of average wind speed reduction across an entire wind farm (due to the deflection of incoming flow, causing part of the flow to bypass the entire farm)†. Such an effect of farm-scale flow reduction has been known to play a key role in the case of tidal-stream turbines in shallow water (e.g., Nishino and Willden 2012, 2013; Garrett and Cummins 2013) but had been considered insignificant for wind turbines for many years, except for the case of an ideal 'infinitely large' wind farm, which has been studied by, e.g., Frandsen (1992), Emeis and Frandsen (1993) and Calaf et al. (2010). In contrast to the traditional 'wake' models (e.g., Lissaman 1979; Jensen 1983; Katić et al. 1986) that describe the reduction of flow behind each turbine, the models that describe the reduction of flow across a very large wind farm in a horizontally-averaged sense (like models for flow through vegetation) are often referred to as 'top-down' models, as discussed in detail by Meneveau (2012). More recently, Stevens et al. (2015, 2016) have proposed a coupled 'wake' and 'top-down' model (called CWBL model) and showed that such a coupled model may predict the statistical (or ensemble-averaged for a given wind direction) performance of a large finite-size wind farm better than traditional wake models. However, CWBL is a pragmatic, engineering-oriented model derived from two existing low-order flow models (rather than directly from the principles of fluid mechanics), meaning that it is inherently subject to limitations due to the underlying low-order flow models. To better understand the true nature of the problem and to provide a new basis for future wind farm modelling at different levels of complexity, it is beneficial to develop a more general 'theory' of wind farm aerodynamics that describes the relationship between the macroscopic flow over an entire farm and the microscopic flow around each turbine without restricting ourselves to specific flow models for each scale. See, e.g., Porté-Agel et al. (2019) for a more comprehensive review of the literature on wind farm modelling.

---

[*] Email address for correspondence: takafumi.nishino@eng.ox.ac.uk

The two-scale momentum theory that we propose in this paper is somewhat similar to the CWBL model of Stevens et al. (2015, 2016) but different in that its aim is to describe a generic relationship between turbine-scale and farm-scale flow problems without specifying the details of flow models at each scale. In particular, we avoid using the logarithmic law explicitly, on which most of the existing top-down models are based. Instead, here we derive our theory directly from the law of momentum conservation, so that the theory may account for the effect of large-scale motions of the atmospheric boundary layer (ABL) in a time-dependent (rather than statistical) manner. This makes it possible to use the present theory to combine, for example, a numerical weather prediction (NWP) model with various types of turbine array models to estimate the wind farm blockage effect for a given large wind farm under given atmospheric (or environmental) conditions. Some of the key concepts employed in the present theory originate from the model of Nishino (2016) proposed for an ideal very large wind farm, which is shown to be derived as a special case from the present theory later in this paper. In the following, we first present the theory in a rigorous manner for a simplified large wind farm situation in Section 2. We then discuss in Section 3 how the theory may be applied in an approximate manner to more realistic large wind farm situations where some of the simplification assumptions employed in the theory are not fully satisfied. We also discuss the limitations and future prospects of the present work in Section 3, followed by conclusions in Section 4 and an Appendix.

## 2. Theory

### *2.1. External momentum balance*

Let us consider a large wind farm over flat terrain or sea surface, as illustrated in figure 1. The horizontal length scale of the farm, $L_F$, is much larger than the thickness of the ABL, $\delta_{\text{ABL}}$, which is typically 1 to 2 km. We consider a short-time averaged flow, i.e., we consider large-scale fluctuations due to changes in atmospheric conditions (with periods of more than about an hour) but not small-scale ones due to turbulence (with periods of typically less than a few minutes). We assume that: (i) many identical wind turbines are arranged regularly over the whole farm area; (ii) the magnitude and vertical profile of 'undisturbed' wind may change in time but they do not vary spatially over the whole farm area at any time, since the horizontal scale of the local atmospheric system driving wind over the farm area is usually much larger than $L_F$; and (iii) the flow over the turbine array (or the internal boundary layer, IBL) is in a fully developed state except for a limited region near the farm edge, i.e., all turbines except for those located near the farm edge have the same flow conditions. In reality, these assumptions may not be fully satisfied and the flow conditions around each turbine may vary over the entire farm. However, the theoretical analysis presented below may still be modified and applied (in an approximate manner) to such a real-world situation, as discussed later in Section 3.1.

---

† Note that this so-called wind farm blockage effect is different from the local blockage effect that increases the power of individual wind turbines placed side-by-side and close to one another (Nishino and Draper 2015), and also different from the global (cross-sectional) blockage effect that increases the power of a turbine or turbines placed in a confined flow passage, such as a closed test section of a wind tunnel (Garrett and Cummins 2007). The increase in turbine power due to the local/global blockage effect is relative to the case with no (or less) confinement of flow, whereas the decrease in farm power due to the wind farm blockage effect is relative to the (hypothetical) case in which the macroscopic flow outside of the farm is not affected by the farm itself. Hence, this wind farm blockage effect may as well be referred to as wind farm induction effect.

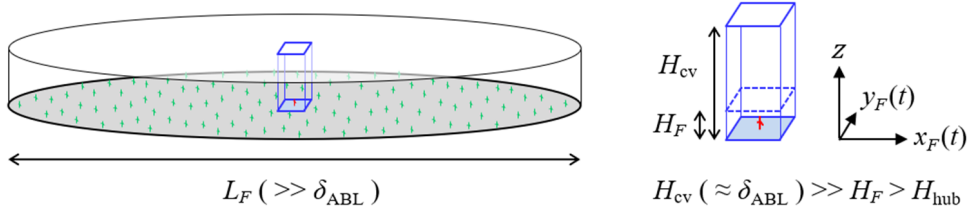

FIGURE 1. Schematic of a large wind farm, representative control volume (CV) within the farm, coordinate system and length scales/dimensions considered.

The above assumptions allow us to make a simplified analysis of farm efficiency by considering a representative, rectangular control volume (CV) containing only one turbine in the middle of the farm (note that this is just for the sake of simplicity; we may also consider a larger CV containing a group of more than one turbine to allow for the existence of periodic flow features with a scale larger than the scale of a single turbine). The CV's horizontal area, $S_{cv}$, corresponds to the farm area per turbine (or group of turbines), whereas its height, $H_{cv}$, is large enough to have a negligibly small shear stress at the top, i.e., $H_{cv} \approx \delta_{ABL}$. The wind direction may change in altitude ($z$) and time ($t$), but the CV's side faces are always aligned to the farm's 'streamwise' direction, $x_F(t)$, defined as the direction of the horizontally averaged flow at the turbine hub-height, $H_{hub}$ (typically about 100m, which is much less than $\delta_{ABL}$). The height of a nominal farm layer, $H_F$, is not required at this stage and will be given later in Section 2.2.

Now we consider the momentum balance for this representative CV. The streamwise momentum equation for a short-time averaged flow is expressed, using the material derivative, $D/Dt$, as

$$\rho \frac{DU}{Dt} = -\frac{\partial p}{\partial x_F} + \left(\frac{\partial \tau_{x_F x_F}}{\partial x_F} + \frac{\partial \tau_{x_F y_F}}{\partial y_F} + \frac{\partial \tau_{x_F z}}{\partial z}\right) + f_{x_F}, \quad (2.1)$$

where $\rho$, $U$ and $p$ are the fluid density, streamwise velocity and pressure, respectively, $x_F$ and $y_F$ the horizontal coordinates (streamwise and lateral directions, respectively, which may change in time but are always perpendicular to each other), $\tau_{ij}$ denotes the stress (mainly the Reynolds stress resulting from the short-time averaging process), and $f_{x_F}$ the body force acting in the streamwise direction per unit volume, including the Coriolis force as described below. The drag due to the turbine may also be considered as part of the body force here, even though this drag is non-zero for only a small fraction of the CV (note that the stress term will implicitly include the dispersive stress if the flow discussed here is a spatially filtered one and does not resolve the spatial inhomogeneity caused by the turbine, but this will not affect the following analysis explicitly). By integrating (2.1) over the CV and noting the assumption that the same flow pattern around each turbine (or each group of turbines) is repeated horizontally over the entire farm (except for the farm edge region), we obtain

$$\int \frac{\partial(\rho U)}{\partial t} dV_{cv} = -(\langle p_{out}\rangle - \langle p_{in}\rangle)\Delta y_F H_{cv} - \langle \tau_w \rangle \Delta x_F \Delta y_F + \int f_{x_F} dV_{cv}, \quad (2.2)$$

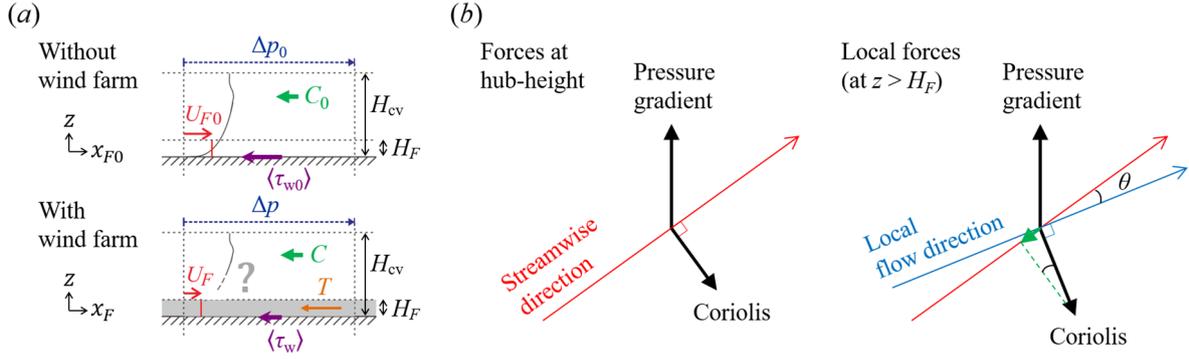

FIGURE 2. Schematic of flows and forces: (*a*) fully developed flows observed in a representative CV without and with wind farm; (*b*) local flow and force vectors at the hub-height (left) and at a higher altitude (right) with the 'streamwise' component of the Coriolis force represented by the green arrow.

where $V_{cv}\ (=S_{cv}H_{cv}=\Delta x_F \Delta y_F H_{cv})$ is the volume of the CV, $\Delta x_F$ and $\Delta y_F$ are the streamwise and lateral lengths of the CV, respectively, $\langle p_{out} \rangle$ and $\langle p_{in} \rangle$ are the pressure averaged over the outlet (downstream) and inlet (upstream) surfaces of the CV, respectively, and $\langle \tau_w \rangle$ is the streamwise shear stress averaged over the bottom surface of the CV. Note that the only shear stress that appears in (2.2) is $\langle \tau_w \rangle$ but this does *not* mean that the effect of mixing inside the CV is ignored. Mixing affects the strength of 'streamwise' Coriolis force described below and thus the momentum balance in (2.2).

Next, we consider the momentum balance given in (2.2) for two different cases: one is with wind farm and the other is without wind farm. For the former case, both Coriolis force and turbine drag contribute to the last term in (2.2). Note that this Coriolis force could be generated not only by the Earth's rotation but also by the change of the streamwise direction itself (especially when it changes rapidly in time) as this causes an additional rotation of the coordinate system, but in the following we ignore this additional rotation effect for simplicity. Although the Coriolis force acts in the direction perpendicular to the local flow direction, this may still affect the farm's streamwise ($x_F$) momentum balance since the local flow direction may change in altitude ($z$) and thus be different from the streamwise direction, as illustrated in figure 2. Hence, the last term in (2.2) can be rewritten as

$$\int f_{x_F}\, dV_{cv} = -T - f_c \int (\rho U \tan\theta)\, dV_{cv}, \quad (2.3)$$

where $T$ is the turbine drag, $f_c$ is the Coriolis parameter ($f_c = 2\Omega \sin\phi$, where $\Omega = 7.292 \times 10^{-5}$ rad/s is the rotation rate of the Earth and $\phi$ is the latitude) and $\theta$ is the angle of local flow direction measured from the streamwise direction ($\theta$ is taken positive in the clockwise direction in figure 2(*b*), looking down from the top of the ABL for the Northern Hemisphere and looking up from the bottom for the Southern Hemisphere). Note that the local velocity in the local flow direction is $U(\cos\theta)^{-1}$ since $U$ is the velocity in the farm's streamwise direction ($x_F$). We can expect that the (horizontally averaged) flow direction does not vary substantially across the thin nominal farm layer of height $H_F$ given later in Section 2.2, but at a higher altitude the flow direction varies and the angle $\theta$ tends to be positive due to the Ekman effect, yielding a component of the Coriolis force opposing to the farm's streamwise direction as shown by the green arrow in figure 2(*b*). By substituting (2.3) into (2.2) and using square brackets to represent volume-averaging over the CV, we obtain

$$\frac{\partial [\rho U]}{\partial t} = \frac{\Delta p}{\Delta x_F} - \frac{T + \langle \tau_w \rangle S_{cv}}{V_{cv}} - f_c[\rho U \tan\theta], \quad (2.4)$$

where $\Delta p\ (=\langle p_{in}\rangle - \langle p_{out}\rangle)$ is the average pressure drop in the streamwise direction across the CV. By repeating the same analysis for the case without the wind farm, we also obtain

$$\frac{\partial [\rho_0 U_0]}{\partial t} = \frac{\Delta p_0}{\Delta x_{F0}} - \frac{\langle \tau_{w0} \rangle S_{cv}}{V_{cv}} - f_c[\rho_0 U_0 \tan\theta_0], \quad (2.5)$$

where the subscript '0' indicates that the variable is for the case without farm. It should be noted that the streamwise direction for the case without farm, $x_{F0}$, may be different from that for the case with farm ($x_F$); hence, for example, $U_0$ is the velocity in $x_{F0}$ and not in $x_F$. By substituting (2.5) into (2.4) for $V_{cv}$, we obtain a combined (non-dimensionalised) momentum equation:

$$\frac{T + \langle \tau_w \rangle S_{cv}}{\langle \tau_{w0} \rangle S_{cv}} = \frac{\frac{\Delta p}{\Delta x_F} - C - \frac{\partial}{\partial t}[\rho U]}{\frac{\Delta p_0}{\Delta x_{F0}} - C_0 - \frac{\partial}{\partial t}[\rho_0 U_0]}, \qquad (2.6)$$

where $C = f_c[\rho U \tan \theta]$ and $C_0 = f_c[\rho_0 U_0 \tan \theta_0]$. The left-hand-side of (2.6) represents the ratio of the streamwise momentum lost by 'total bottom resistance' (including turbine drag) for the case with farm ($T + \langle \tau_w \rangle S_{cv}$) to that for the case without farm ($\langle \tau_{w0} \rangle S_{cv}$), whereas the right-hand-side shows the ratio of the streamwise momentum available to the total bottom resistance for the case with farm to that for the case without farm. Hence, for convenience, we introduce a new parameter called the momentum availability factor, $M$, to denote the right-hand-side of (2.6), i.e.,

$$M \equiv \frac{\frac{\Delta p}{\Delta x_F} - C - \frac{\partial}{\partial t}[\rho U]}{\frac{\Delta p_0}{\Delta x_{F0}} - C_0 - \frac{\partial}{\partial t}[\rho_0 U_0]}. \qquad (2.7)$$

As will be discussed later in Section 2.4, $M$ is a parameter depending on several external (farm-scale) conditions but can be modelled numerically using a NWP model. Specifically, $M$ will be modelled as a function of the farm wind-speed reduction factor that is defined below.

## 2.2. Farm wind-speed reduction factor

Now we define the 'farm-average' wind speed, $U_F$, by introducing a thin 'nominal farm layer' of height $H_F$ as depicted earlier in figure 2. The purpose of defining $U_F$, and thus the farm wind-speed reduction factor, $\beta \equiv U_F/U_{F0}$, is not only for the modelling of $M$ but also for the left-hand-side of (2.6), i.e., change of momentum loss due to the turbine drag and shear stress on the bottom surface. Eventually, both left- and right-hand-sides of (2.6) will be functions of $\beta$, allowing us to calculate $\beta$ for a given set of external (farm-scale) and internal (turbine-scale) conditions. The role of $\beta$ is thus, essentially, to provide a link between the external problem described in Section 2.1 (which is a time-dependent problem considering large-scale motions of the ABL to assess the momentum available to the total bottom resistance at a certain time) and the internal problem described later in Section 2.3 (which is a quasi-steady problem giving the breakdown of the total bottom resistance into the turbine drag and the bottom shear stress). It is worth noting that the role of $\beta$ (or more precisely, $1 - \beta$, which may be referred to as 'farm induction factor') is analogous to that of 'array-scale induction factor' introduced by Nishino and Willden (2012) for their two-scale modelling of tidal turbine arrays.

There are a few possible ways to define $H_F$ and $U_F$, but here we employ the approach proposed by Nishino (2016); see also Section 2.1 of Nishino and Hunter (2018) for details. This approach defines $H_F$ based on a 'natural' wind profile, $\overline{U_0}(z)$, which is a long-time-average of the streamwise velocity profile for the case without farm, $U_0(z,t)$. Specifically, $H_F$ is defined as the farm-layer height with which the value of $\overline{U_0}$ averaged over the farm layer agrees with that averaged over the turbine's rotor swept area, i.e.,

$$\frac{\int_0^{H_F} \overline{U_0} dz}{H_F} = \frac{\int \overline{U_0} dA}{A}, \qquad (2.8)$$

where $A$ is the rotor swept area. A typical value of $H_F$ is between $2H_{hub}$ and $3H_{hub}$ depending on the turbine design and the ABL profile. With the above definition of $H_F$, now $U_F$ and $U_{F0}$ are defined as

$$U_F \equiv \frac{\int \left( \int_0^{H_F} U dz \right) dS_{cv}}{H_F S_{cv}} \quad \text{and} \quad U_{F0} \equiv \frac{\int_0^{H_F} U_0 dz}{H_F}. \qquad (2.9a \text{ and } b)$$

This allows us to introduce a 'local' or 'internal' thrust coefficient of the turbine, $C_T^*$, defined using $U_F$ as the reference wind speed, i.e.,

$$C_T^* \equiv \frac{T}{\frac{1}{2}\rho_F U_F^2 A}, \qquad (2.10)$$

where $\rho_F$ is the fluid density averaged over the farm layer for the case with farm, but this should be almost identical to that for the case without farm, $\rho_{F0}$. Note that here we assume that the turbine drag is all due to the rotor thrust (and this is why the reference area for $C_T^*$ is the rotor swept area, $A$) but the turbine's support-structure drag may also be considered in a similar manner if necessary (Ma and Nishino 2018). In addition to $C_T^*$, we also introduce a bottom friction exponent, $\gamma$, which is defined as

$$\gamma \equiv \log_\beta(\langle \tau_w \rangle / \langle \tau_{w0} \rangle), \qquad (2.11)$$

where $\beta \equiv U_F/U_{F0}$. As will be discussed later in Section 2.3, $C_T^*$ and $\gamma$ are parameters depending on several internal (turbine-scale) conditions; the former gives a relationship between $T$ and $U_F$, whereas the latter gives a relationship between $\langle \tau_w \rangle$ and $U_F$. By substituting (2.7), (2.10) and (2.11) into (2.6), and assuming $\rho_F = \rho_{F0}$, the momentum equation (2.6) can be transformed into

$$C_T^* \frac{\lambda}{C_{f0}} \beta^2 + \beta^\gamma = M, \qquad (2.12)$$

where $\lambda \equiv A/S_{cv}$ is the farm density (or array density) and $C_{f0}$ is a bottom friction coefficient for the case without farm, defined as

$$C_{f0} \equiv \frac{\langle \tau_{w0} \rangle}{\frac{1}{2}\rho_{F0} U_{F0}^2}. \qquad (2.13)$$

The parameter $\lambda/C_{f0}$ in (2.12) is referred to as the effective farm density (Nishino 2016, Nishino and Hunter 2018). A typical range of $\lambda/C_{f0}$ is between 1 and 10, depending on the roughness of land/sea surface as well as on the inter-turbine spacing. The first and second terms of (2.12) describe 'relative' momentum losses due to the turbine drag and the bottom shear stress, respectively (relative to the natural momentum loss for the case without farm). Note that the transformed momentum equation (2.12) is still almost identical to the original momentum equation (2.6) since the only approximation made during the transformation is that for the farm-average fluid density ($\rho_F = \rho_{F0}$). Hence (2.12) should be almost exactly satisfied if the values of $C_T^*$, $\gamma$ and $M$ are all accurate (for a given 'fully developed' farm at a given farm location and time).

Before discussing how to model $C_T^*$ and $\gamma$, it should be noted that the height of the farm layer, $H_F$, may be defined differently from the above. For example, we may define $H_F$ simply as a fixed height, e.g., $H_F = 2.5 H_{\text{hub}}$, instead of using (2.8) which requires the natural wind profile. Differences in the definition of $H_F$ will affect the values of $C_{f0}$, $C_T^*$ and $\gamma$ (and also the value of 'momentum response factor,' $\zeta$, which will be introduced later for the modelling of $M$); however, the momentum equation (2.12) will be unchanged by the definition of $H_F$. In other words, (2.12) is valid (and can therefore serve as the condition for coupling between the internal and external problems) as long as the same value of $H_F$ is used in the modelling of both internal and external problems. The theory is expected to remain physically reasonable if (i) $H_F$ is much smaller than the ABL thickness (i.e., $H_F \ll \delta_{\text{ABL}}$) and (ii) $H_F$ is large enough to include the region where the flow is most strongly affected by the turbines (i.e., $H_F > H_{\text{hub}} + R$, where $R$ is the rotor radius). The main advantage of employing (2.8) is that it gives a link between the natural wind speed averaged over the farm layer and that averaged over the rotor swept area, thus simplifying the relationship between the power coefficient of the turbine and a (non-dimensional) power density of the farm that will be given later in Section 2.5.

### 2.3. Internal momentum balance

Now we briefly discuss the modelling of $C_T^*$ and $\gamma$, which together describe the internal momentum balance in the left-hand-side of (2.12), i.e., balance between the momentum lost by the turbine drag and that lost by the bottom shear stress, for a given $\beta$ (remember that we need $M$ as well as $C_T^*$ and $\gamma$ to obtain $\beta$). In general, $C_T^*$ and $\gamma$ may depend on several internal (turbine-scale) conditions, such as the design and operating conditions of the turbines and their array configuration, as well as on the conditions of wind over the turbine array, including its speed, direction and turbulence characteristics. Some external (farm-scale) conditions, such as the size and location of the farm (and of nearby farms if they exist) and the type of local atmospheric system that drives wind over the farm, may also affect $C_T^*$ and $\gamma$ indirectly because these conditions may affect the conditions of wind over the turbine array (as will be discussed further in Section 2.4). However, it is impractical to analyse the influence of all internal and external conditions simultaneously. This is why the present theory splits the problem into the internal and external problems; $C_T^*$ and $\gamma$ are modelled in the former and $M$ in the latter.

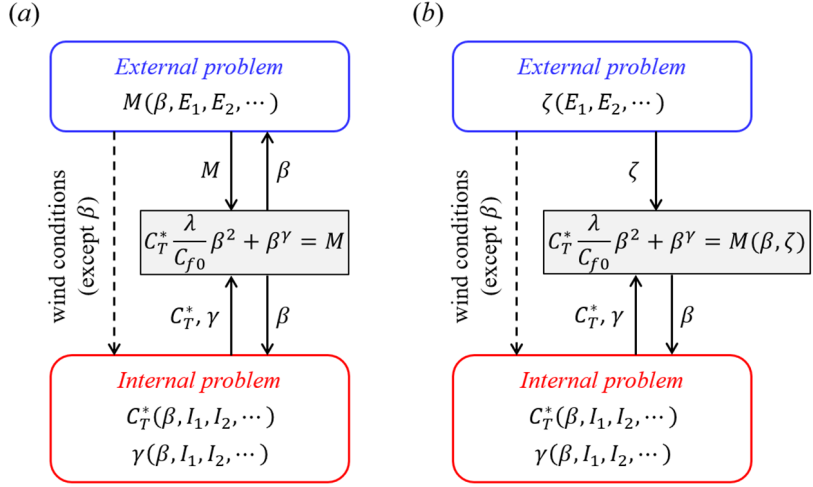

FIGURE 3. Relationship between the external and internal problems: (*a*) general case, where the two problems are loosely coupled via *β*; (*b*) simplified case, where the external problem is decoupled from the internal problem. $E_i$ and $I_i$ represent external and internal conditions, respectively.

In the internal problem, we do not consider the influence of external conditions explicitly, although we may still prescribe various conditions of wind over the turbine array (that are in reality influenced by some external conditions) to assess their effects on $C_T^*$ and $\gamma$. We also do not consider any direct effect of large-scale fluctuations that are considered in the external problem (with typical periods of more than an hour) since their time scale is much larger than that of the flow around each turbine. This basically means that the internal problem is considered as a quasi-steady problem[‡], i.e., large-scale fluctuations may affect the internal problem (and thus $C_T^*$ and $\gamma$) only indirectly through *β* and the prescribed wind profiles (that may result from large-scale fluctuations). The only flow condition that depends explicitly on the two-way interaction between the internal and external problems is the magnitude of wind, which is readily determined by the value of *β* that is obtained from (2.12). Hence, the internal problem is only loosely coupled to the external problem (and this may even be decoupled by introducing a further simplification, as will be described below and illustrated in figure 3).

The internal problem may be modelled either numerically or analytically to obtain $C_T^*$ and $\gamma$. If we employ a high-fidelity numerical model, such as Large-Eddy Simulations (LES) of ABL flow over a periodic array of turbines represented using an actuator line method (e.g., Lu and Porté-Agel 2011), we would obtain highly accurate values of $C_T^*$ and $\gamma$ for a specific case. However, such high-fidelity simulations require large computational resources and thus cannot be employed to assess the effects of a wide range of internal conditions. If we temporarily ignore the effects of turbine rotor's details and focus on the effects of other internal conditions, then LES combined with an actuator disc model or a porous disc model (e.g., Calaf et al. 2010) would be an alternative option. For example, Ghaisas et al. (2017) and Dunstan et al. (2018) have conducted such LES to investigate the effects of turbine array configuration, ground (or sea-surface) roughness and atmospheric stability condition on $C_T^*$ and $\gamma$. The benefits of employing a simple actuator disc model are not only that the computational cost is reduced but also that the internal problem becomes insensitive to the value of $\beta$ (as the performance characteristics of an actuator disc do not depend on the absolute value of wind speed, unlike a more detailed model that takes into account the dependence of turbine's characteristics on the wind speed, e.g., whether the wind speed is above or below the rated wind speed), making it possible to assess the effects of other internal conditions on $C_T^*$ and $\gamma$ independently from the external problem.

---

[‡] However, we may consider small-scale fluctuations due to turbulence explicitly in the internal problem. In this case, the flow modelled for the internal problem needs to be short-time averaged to obtain the values of $C_T^*$ and $\gamma$ that represent the local thrust coefficient and bottom friction exponent at a 'given time' from the viewpoint of the (longer-time-scale) external problem.

The above LES studies by Ghaisas et al. (2017) and Dunstan et al. (2018), and also a similar study by Zapata et al. (2017) using Reynolds-averaged Navier-Stokes (RANS) simulations, suggest that $C_T^*$ may be predicted fairly well for a range of internal conditions using a simple analytical model. This analytical model, proposed first by Nishino (2016) in conjunction with the definition of the nominal farm layer discussed earlier, gives $C_T^*$ simply as a function of wind speed reduction at the rotor plane (using an analogy with the classical actuator disc theory for an isolated wind turbine) as

$$C_T^* = 4\alpha(1 - \alpha), \qquad (2.14)$$

where $\alpha \equiv U_T/U_F$ is a 'local' or 'internal' (turbine-scale) wind speed reduction factor, and $U_T$ is the streamwise velocity averaged over the turbine's rotor swept area, i.e.,

$$U_T \equiv \frac{\int U dA}{A} . \qquad (2.15)$$

This is arguably the simplest possible model of $C_T^*$, which takes into account the effect of local wind speed reduction (or turbine resistance) only and does not consider any other conditions, such as array configuration, wind direction and wind profile, explicitly. Nevertheless, unless neighbouring turbines are aligned perfectly with wind direction to cause a significant level of direct wake interference, the $C_T^*$ value calculated from (2.14) tends to agree fairly well with the true $C_T^*$ value (with a typical error of less than 10%) for a realistic range of inter-turbine spacing (Nishino 2016, Zapata et al. 2017) and for various wind profiles induced by different atmospheric stability conditions (Dunstan et al. 2018). A further investigation into the validity of (2.14) is shown in Appendix A. Apart from its simplicity, a major advantage of this approach using an analogy with the actuator disc theory is that it can be easily combined with the blade-element theory to assess the effects of turbine rotor design and operating conditions on $C_T^*$; see Nishino and Hunter (2018) for further details. If we employ (2.14) as the model of $C_T^*$ and substitute it into the momentum equation (2.12), we obtain

$$4\alpha(1-\alpha)\frac{\lambda}{C_{f0}}\beta^2 + \beta^\gamma = M . \qquad (2.16)$$

Note that, if we assume $M = 1$ (i.e., if the momentum available to the total bottom resistance does not change between the cases with and without wind farm), this equation (2.16) becomes identical to the original two-scale momentum model of Nishino (2016) (which predicts an upper limit of power generation from an ideal, infinitely large wind farm with a fixed amount of momentum per unit area supplied by an ideal, infinitely large atmospheric system). In other words, (2.16) can be seen as a generalised version of the two-scale momentum model of Nishino (2016).

For the modelling of $\gamma$, the LES study by Ghaisas et al. (2017) suggests that this parameter may be modelled, for the case of a neutral ABL, as a function of the turbine resistance coefficient, $C_T' = C_T^*/\alpha^2$, multiplied by the array density. However, further investigations are required in the future to develop a model of $\gamma$ for a wider range of internal conditions. Nevertheless, as discussed by Nishino (2016), the value of $\gamma$ is expected to be close to and less than 2 in most cases, since the value of a 'local' bottom friction coefficient, $C_f \equiv \langle \tau_w \rangle / \frac{1}{2}\rho_F U_F^2$, tends to be larger than its undisturbed value, $C_{f0}$ (due to the effects of turbines increasing turbulence intensity and local flow inhomogeneity within the nominal farm layer). The aforementioned LES studies by Ghaisas et al. (2017) and Dunstan et al. (2018) also suggest that the value of $\gamma$ is in the range between 1.5 and 2 for most cases and, as will be shown later in Section 2.5, the farm performance predicted using the present theory is not very sensitive to the value of $\gamma$ in this range. Hence, unless sufficient data are available, it is acceptable to employ a fixed $\gamma$ value, for example, $\gamma = 2$ (which gives $C_f = C_{f0}$, i.e., $\langle \tau_w \rangle$ varies with $U_F^2$) as a first order approximation.

## 2.4. Momentum availability factor

Now we return to the external problem for the modelling of the right-hand-side of (2.12), namely the farm's momentum availability factor, $M$. In the external problem we do not consider the effect of any internal conditions explicitly; hence, internal conditions may affect $M$ only indirectly through $\beta$, as illustrated earlier in figure 3a. This simplification allows us to model the external problem (and thus $M$) numerically without resolving any details of flow around each turbine. For example, we may use a regional NWP model with a large wind farm represented simply by an area of increased bottom roughness to assess the effect of large-scale motions of the atmosphere on $M$. The assumption here is that such a simple farm model (that does not resolve individual turbines) can still predict the level of Reynolds stress (averaged over the farm layer) reasonably well for a given $\beta$, so that the macroscopic flow around the entire farm (especially the rate of turbulent mixing downstream of the entire farm) is predicted correctly for a given $\beta$. In reality, the Reynolds stress level and thus the macroscopic flow characteristics may change with some internal conditions, such as the array configuration, even for a fixed value of $\beta$. To account for such secondary effects of internal conditions separately from $\beta$, we would need to employ a more sophisticated farm model that yields a correct level of Reynolds stress for a given set of internal conditions; see, e.g., Fitch et al. (2013) and Abkar and Porté-Agel (2015).

To obtain the value of $M$ numerically using a NWP model, we need to conduct 'twin' simulations, i.e., two simulations under identical initial and boundary conditions except that one is with farm and the other is without farm. Since $M$ depends on $\beta$ (and $\beta$ depends on the internal problem), in general, we need to conduct NWP simulations several times (with varying the farm resistance) iteratively in conjunction with an internal flow model to find a converged value of $\beta$ for a given farm situation (as in figure 3a). However, we may be able to reduce the number of required NWP simulations if we can develop an approximate model of $M$ as a function of $\beta$ and an environment-dependent parameter that does not depend on $\beta$. One example of such a model is a linear approximation model given by

$$M = 1 + \zeta(1 - \beta), \tag{2.17}$$

where $\zeta$ is a non-dimensional parameter, which we refer to as 'momentum response factor' since this describes how the momentum available to the total bottom resistance responds to the change of farm-average wind speed. Although this is a very simple model, a recent numerical study of pressure-driven boundary-layer flow over a large staggered array of actuator discs by Nishino (2018) shows that this linear approximation works well for a practical range of $\beta$ (between 1 and 0.8) with the value of $\zeta$ depending on the roughness length of the land/sea surface around the farm area but not depending on $\beta$. The basic trend is that $M$ becomes larger than 1 (i.e., the momentum available to the total bottom resistance becomes larger than that for the case without farm) as $\beta$ decreases from 1. As discussed by Nishino (2018) this is essentially because an additional pressure difference is induced across the farm area by the resistance caused by the farm itself. The amount of this farm-induced pressure difference depends on the characteristics of macroscopic flow around the entire farm (i.e., how easily or not so easily the flow can bypass the entire farm), which explains why the response factor $\zeta$ depends on the level of land/sea surface roughness. Although the numerical study by Nishino (2018) is for a special case where the acceleration/deceleration of the ABL and the Coriolis force are neglected, an ongoing study using a NWP model with a large circular patch of increased bottom roughness to represent a large offshore wind farm (Ma et al., unpublished) suggests that the linear model given by (2.17) is approximately valid for more realistic unsteady cases as well (with $\zeta$ depending on time).

A major advantage of employing an approximation model of $M$, such as (2.17), is that the external problem can be decoupled from the internal problem, as described in figure 3b. This will allow us to solve the external problem to assess the response characteristics of the ABL for a given farm location (represented by the value of $\zeta$ in this example) separately from, and even before solving, the internal problem. This means that we may evaluate the potential of a given wind farm site not only from the characteristics of wind naturally available at the site but also from its response characteristics (which determine how significant the reduction of farm-average wind speed tends to be at that site) obtained from an independent external flow model.

## 2.5. Power coefficient and power density

Finally, we define the power coefficient of the turbine and a non-dimensional power density of the farm, both of which describe the efficiency of power generation at a given time (from the viewpoint of the time-dependent external problem). The power coefficient, $C_P$, may be defined as

$$C_P \equiv \frac{P}{\frac{1}{2}\rho_{T0}U_{T0}^3 A} = \frac{P}{\frac{1}{2}\rho_{F0}U_{F0}^3 A}\sigma_1 \ , \tag{2.18}$$

where $P$ is the turbine power, $\rho_{T0}$ and $U_{T0}$ are the fluid density and streamwise velocity, respectively, averaged over the turbine rotor swept area (for the case without farm), and $\sigma_1$ is a conversion factor given by

$$\sigma_1 = \frac{\rho_{F0}U_{F0}^3}{\rho_{T0}U_{T0}^3} \ . \tag{2.19}$$

Note that $C_P$ in (2.18) represents the ratio of the (short-time average) turbine power to the (long-time average) power of natural wind passing through the turbine rotor swept area. If we introduce a 'local' or 'internal' power coefficient, $C_P^*$, in a similar manner to $C_T^*$ defined earlier in (2.10), i.e.,

$$C_P^* \equiv \frac{P}{\frac{1}{2}\rho_F U_F^3 A}, \tag{2.20}$$

and assume $\rho_F = \rho_{F0}$ as before, then we obtain

$$\frac{C_P}{\sigma_1} = C_P^* \beta^3 \ . \tag{2.21}$$

Similarly to $C_T^*$, $C_P^*$ is a parameter obtained from the internal problem (where the relationship between $C_T^*$ and $C_P^*$ depends, in general, on many internal conditions including the details of turbine design and operating conditions). We may also define the non-dimensional power density, $\eta$, as

$$\eta \equiv \frac{P}{\langle \tau_{w0} \rangle U_{F0} S_{\text{cv}}} = \frac{P}{\langle \tau_{w0} \rangle U_{F0} S_{\text{cv}}}\sigma_2 \ , \tag{2.22}$$

where $\sigma_2$ is another conversion factor given by

$$\sigma_2 = \frac{\langle \tau_{w0} \rangle U_{F0}}{\langle \tau_{w0} \rangle U_{F0}} \ . \tag{2.23}$$

Note that $\eta$ in (2.22) represents the ratio of the (short-time average) turbine power to the (long-time average) power of wind that is naturally dissipated due to the land/sea surface friction over the farm area per turbine. Noting the definition of $C_{f0}$ given earlier in (2.13) and $\lambda = A/S_{\text{cv}}$, we can derive the general relationship between $\eta$ and $C_P$ as

$$\frac{\eta}{\sigma_2} = \frac{C_P}{\sigma_1}\frac{\lambda}{C_{f0}} \ . \tag{2.24}$$

For special cases where the unsteadiness of the flow is ignored and the fluid density is assumed to be constant across the farm layer, we obtain $\sigma_1 = \sigma_2 = 1$ (noting that (2.8) gives $U_{F0} = U_{T0}$ for such a case) and hence (2.24) returns to $\eta = C_P \lambda / C_{f0}$ as in the model of Nishino (2016).

Although the present theory supposes that the turbine power, and thus the efficiency, are obtained from an arbitrary combination of (usually numerical) flow models, it is still possible and meaningful to calculate the efficiency using the simple analytical/approximation models given earlier in Sections 2.3 and 2.4. Specifically, if we employ (2.14) and (2.17) to model $C_T^*$ and $M$ in (2.12), respectively, we can solve (2.12) to obtain $\beta$ as a function of $\alpha$ for a given set of input parameters: $\gamma$, $\zeta$, $\lambda$ and $C_{f0}$. As the actuator disc concept is used to derive (2.14), we may also consider $P = TU_T$ and $C_P^* = C_T^* \alpha$; hence, $C_P/\sigma_1$ (which represents the turbine power relative to the power of undisturbed wind available at that time, not the long-time-averaged power) now becomes a function of $\alpha$ and $\beta$ only, i.e.,

$$\frac{C_P}{\sigma_1} = \frac{TU_T}{\frac{1}{2}\rho_{F0} U_{F0}^3 A} = 4\alpha^2 (1-\alpha)\beta^3 \ . \tag{2.25}$$

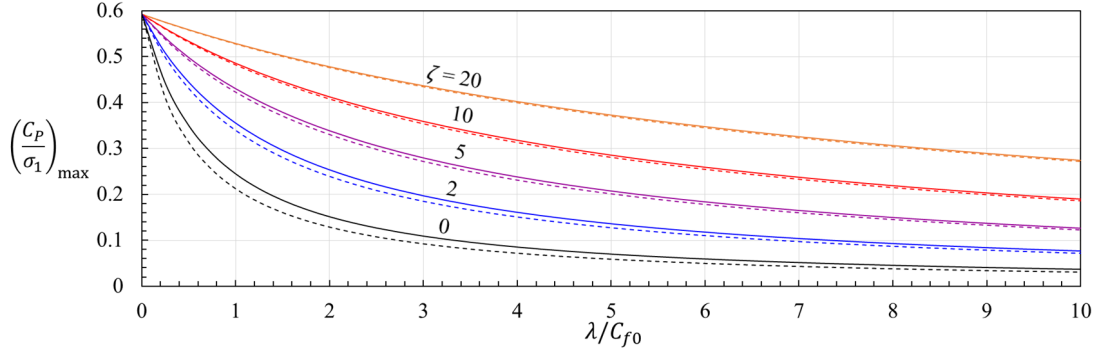

FIGURE 4. Maximum efficiency of a turbine located in a 'fully developed' part of a large wind farm plotted against the effective farm density, predicted by the two-scale momentum theory (2.12) with simplified models of $C_T^*$ (2.14) and $M$ (2.17) (solid lines: $\gamma = 2$; dashed lines: $\gamma = 1.5$).

Therefore, for a given set of input parameters ($\gamma$, $\zeta$, $\lambda$ and $C_{f0}$), we can obtain $\beta$ and then $C_P/\sigma_1$ as a function of $\alpha$. Figure 4 shows the maximum value of $C_P/\sigma_1$ (obtained by varying $\alpha$) plotted against the effective farm density, $\lambda/C_{f0}$, for selected values of $\gamma$ and $\zeta$. As can be seen from the figure, the maximum efficiency decreases from the well-known 'Betz-limit' of 16/27 ($\approx 0.593$) to a lower value as $\lambda/C_{f0}$ increases from zero to a higher value. While the effect of $\gamma$ is relatively minor for a practical range of $\gamma$ (between 1.5 and 2 as noted earlier in Section 2.3), the effect of $\zeta$ seems more significant. Although a typical range of $\zeta$ is still unknown and an extensive numerical study will be needed in the future to assess $\zeta$ under various external conditions, the aforementioned numerical study by Nishino (2018) suggests that, for the case of a steady pressure-driven flow, the value of $\zeta$ may be around 5 to 10 depending on the level of land/sea surface roughness. It is worth noting that $\zeta$ tends to increase as the surface roughness decreases; however, $\lambda/C_{f0}$ also increases as the surface roughness decreases (as $C_{f0}$ decreases) and as a result, for a given array of turbines, the maximum efficiency $(C_P/\sigma_1)_{max}$ still tends to decrease with the surface roughness (Nishino 2018).

The decrease in the maximum efficiency predicted here is essentially due to the reduction of wind speed across the entire farm. This can be seen from figure 5, which shows the values of $\alpha$ and $\beta$ that yield the maximum efficiency, namely the optimal turbine-scale and farm-scale wind-speed reduction factors, $\alpha_{opt}$ and $\beta_{opt}$. These optimal values also depend significantly on $\zeta$ and less significantly on $\gamma$, but the general trend is that $\beta_{opt}$ decreases and $\alpha_{opt}$ increases as $\lambda/C_{f0}$ increases. When $\lambda/C_{f0} = 0$, the value of $\beta$ is always 1 and hence $\alpha_{opt}$ is 2/3 ($\approx 0.667$) to maximise the value of $4\alpha^2(1-\alpha)\beta^3$. When $\lambda/C_{f0} > 0$, however, $\beta$ tends to decrease as $C_T^*$ increases; hence $\alpha_{opt}$ becomes higher than 2/3 to reduce $C_T^*$ and eventually maximise the value of $4\alpha^2(1-\alpha)\beta^3$. This basically means that the optimal resistance of a turbine (or an actuator disc) in a large wind farm is lower than that of an isolated one because of the effect of reduced $\beta$, lowering the maximum efficiency from the Betz limit of 16/27 (via the reduction of both $4\alpha^2(1-\alpha)$ and $\beta^3$).

It should be remembered that the results shown in figures 4 and 5 rely on the simple actuator disc concept for the modelling of $C_T^*$ (2.14) and the relationship between $C_T^*$ and $C_P^*$ (i.e., $C_P^* = C_T^* \alpha$). In reality, both $C_T^*$ and $C_P^*$ (and their relationship) depend on the details of turbine design and operating conditions (Nishino and Hunter 2018) as well as on other internal conditions (e.g., the array layout); hence the results will be more complicated. Nevertheless, this simple example using the actuator disc concept demonstrates how the present theory can be used to determine the farm wind-speed reduction factor, and thus the efficiency of power generation, from the (modelled) solutions of both internal and external problems in a combined manner.

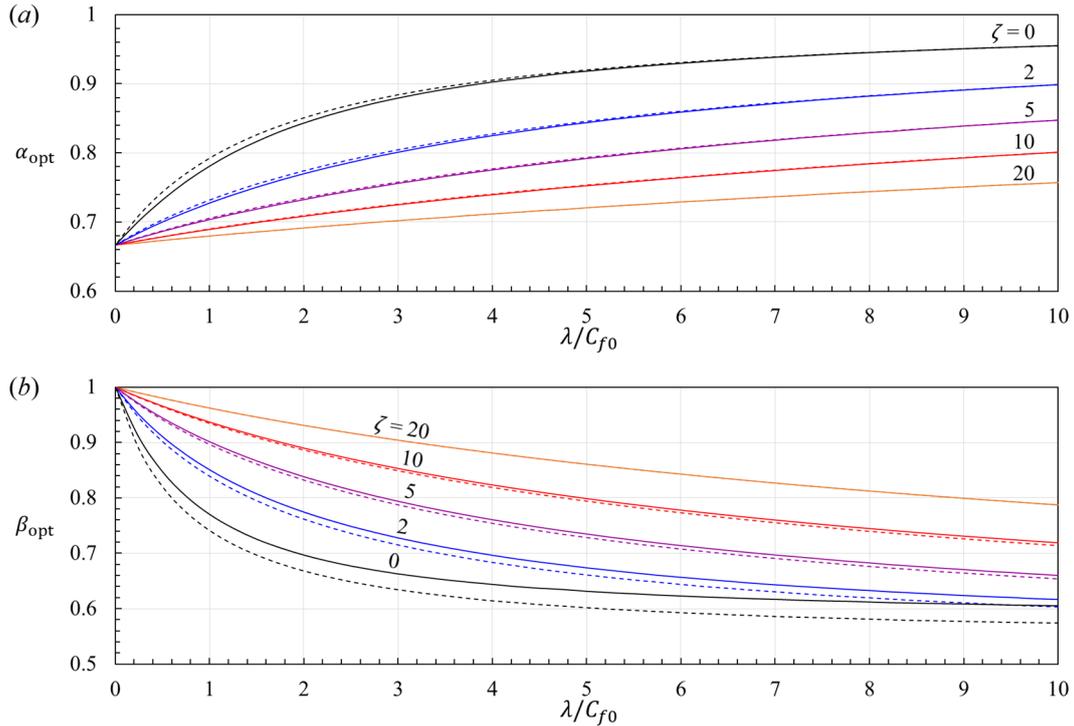

FIGURE 5. The values of (*a*) turbine-scale and (*b*) farm-scale wind-speed reduction factors that yield the maximum efficiency presented in figure 4, predicted by the two-scale momentum theory (2.12) with simplified models of $C_T^*$ (2.14) and $M$ (2.17) (solid lines: $\gamma = 2$; dashed lines: $\gamma = 1.5$).

## 3. Discussion

### 3.1. Horizontal variations across the farm

The theory presented above is based on the assumption that the flow over the turbine array is in a fully developed state, i.e., the same local flow pattern around each turbine (or each group of turbines) is repeated over the entire farm (except for the farm edge region). In reality, however, the flow over the array may not be fully developed since, for example, the internal boundary layer generated by the wind farm may not merge quickly with the external ABL (and the merged boundary layer may also take a long distance to reach a new equilibrium state) depending on atmospheric stability conditions (Wu and Porté-Agel 2017). This assumption may also be violated simply due to an irregular turbine arrangement, variation of turbine operating conditions across the farm and/or inhomogeneity of the natural atmospheric flow over the farm area. In such a real-world wind farm problem with horizontal variations of local flow conditions, the present theory may still be employed to help analyse the farm efficiency but only in an approximate manner. Specifically, the theory (with minor modifications as described below) still allows us to couple an external flow model, which captures horizontal variation of the natural atmospheric flow over the farm area but ignores or highly simplifies variations caused by turbine-scale details, and an internal flow model, which captures variations caused by turbine-scale details but ignores or highly simplifies the variation of the natural atmospheric flow; and thus predict a farm-average value of the farm wind-speed reduction factor, $\hat{\beta}$.

The modifications required to the theory are as follows. First, we consider the external momentum balance not for the representative CV in the middle of the farm as discussed earlier in Section 2, but for a much larger CV that contains the flow over the entire farm area, such as the cylindrical volume depicted in figure 1 for a circular wind farm case. Second, we introduce a single (farm-average) farm layer height, $\widehat{H_F}$, either using (2.8) with replacing $\overline{U_0}$ with its horizontal average over the entire farm, or using an arbitrary definition such as $\widehat{H_F} = 2.5 H_{\text{hub}}$ (as discussed earlier in Section 2.2)[§]. Then we can still derive a momentum equation for two-scale coupling in the same form as (2.12) but with all variables replaced by corresponding variables defined for the entire farm. Specifically, the left-hand-side of (2.12) will represent the internal momentum balance for the entire farm if $\beta$, $\lambda$, $C_{f0}$, $C_T^*$ and $\gamma$ are replaced by the following 'farm-average' counterparts:

$$\hat{\beta} \equiv \frac{\widehat{U_F}}{\widehat{U_{F0}}}, \tag{3.1}$$

$$\hat{\lambda} \equiv \frac{NA}{S_F}, \tag{3.2}$$

$$\widehat{C_{f0}} \equiv \frac{\widehat{\tau_{w0}}}{\frac{1}{2}\widehat{\rho_{F0}}\widehat{U_{F0}}^2}, \tag{3.3}$$

$$\widehat{C_T^*} \equiv \frac{\frac{1}{N}\sum_{i=1}^{N} T_i}{\frac{1}{2}\widehat{\rho_F}\widehat{U_F}^2 A}, \tag{3.4}$$

$$\hat{\gamma} \equiv \log_{\hat{\beta}}(\widehat{\tau_w}/\widehat{\tau_{w0}}), \tag{3.5}$$

where a hat denotes a farm-average value (i.e., value averaged horizontally over the farm area), $N$ is the number of turbines in the farm, $T_i$ is the turbine drag for the $i$-th turbine, and $S_F$ is the farm area. Note that the velocities and shear stresses are again for the farm's 'streamwise' direction defined as the direction of the horizontally averaged flow at $H_{\text{hub}}$ (for each of the cases with and without farm). Meanwhile, the right-hand-side of (2.12), or the momentum availability factor, $M$, will be in a more complicated form than that given earlier for the fully developed case in (2.7), since now we need to consider the effect of (generally non-zero) net momentum transfer through the side (and also top, unless the ABL thickness is constant over the farm area) surfaces of the CV in addition to the effects of the pressure gradient, local acceleration/deceleration and the Coriolis force. While the original $M$ in (2.7) is relatively simple and may perhaps be modelled analytically in a future study, the one for the general case considered here, namely $\widehat{M}$, is more difficult to be modelled analytically. Nevertheless, this can still be obtained using a numerical model in the same manner as discussed in Section 2.4.

The basic procedure for calculating $\hat{\beta}$ for a given wind farm (under given atmospheric conditions) would therefore be as follows. First, the internal problem is modelled to calculate $\widehat{C_{f0}}$, $\widehat{C_T^*}$ and $\hat{\gamma}$ for a given external flow condition (typically by fixing either velocity or pressure outside the farm). If the internal problem is modelled using a three-dimensional (Navier-Stokes-based) numerical model, all these parameters are obtained directly from the model; however, if a low-order 'engineering' model (such as those relying on the traditional 'wake' models mentioned earlier) is employed, only $\widehat{C_T^*}$ may be obtained from the model and $\widehat{C_{f0}}$ and $\hat{\gamma}$ may need to be estimated empirically. Second, the external problem is modelled to calculate $\widehat{M}$ for a given total bottom resistance (for example, by prescribing an increased bottom roughness to represent the whole farm). Third, the obtained values of $\widehat{C_{f0}}$, $\widehat{C_T^*}$, $\hat{\gamma}$ and $\widehat{M}$ (together with the value of $\hat{\lambda}$ determined from the array configuration) are substituted into (2.12) to calculate $\hat{\beta}$. Then both internal and external problems are updated such that the value of $\hat{\beta}$ realised in each problem agrees with that calculated from (2.12) (by adjusting the previously given conditions, such as fixed external velocity or pressure for the internal problem and the level of increased bottom roughness for the external problem) to obtain updated values of $\widehat{C_T^*}$, $\hat{\gamma}$ and $\widehat{M}$; and the same process is repeated as in figure 3$a$ until a converged value of $\hat{\beta}$ is obtained. Eventually, the power generated by the farm (taking into account the loss due to the wind farm blockage effect) can be obtained from the internal problem with the correct (converged) value of $\hat{\beta}$.

---

[§] Such a simple definition of $\widehat{H_F}$ may be preferable in a real-world problem, where the natural wind profile may have a local maximum or maxima around the turbine hub-height (see, e.g., Kettle (2014)) and this may prevent us from obtaining an appropriate farm-layer height from (2.8). It is also worth noting that traditional 'top-down' models (employing the logarithmic law to represent the natural wind profile) are not directly applicable to such cases with a local maximum or maxima in the profile, whereas the present theory (employing the nominal farm layer to define the farm-average wind speed) is still valid and applicable to such cases.

Finally, it should be noted that the amount of power loss due to the wind farm blockage effect can be calculated explicitly by subtracting the final prediction of farm power (obtained using the correct value of $\hat{\beta}$) from an initial (wrong) prediction using a fixed wind speed upstream of the farm. Since the farm power is often approximately proportional to $\hat{\beta}^3$, we may consider $\hat{\beta}_*^3 - \hat{\beta}^3$ as an indicator of the significance of wind farm blockage effect, where $\hat{\beta}_*$ denotes the (farm-average) farm wind-speed reduction factor for the case with a fixed wind speed upstream of the farm.

### 3.2. Limitations and future prospects

The main feature of the two-scale momentum theory is that, as 'momentum' in its name implies, it describes the relationship between the external and internal problems only in terms of the momentum balance through the farm wind-speed reduction factor $\beta$. In other words, the theory does not provide any specific details on how the two problems should be coupled regarding the flow conditions other than $\beta$, such as the wind direction and vertical profiles of wind and turbulence. While the advantage of this theory is its generality or compatibility with many different types of flow models that may be employed at each scale, the details of the flow conditions (other than $\beta$) given to the internal problem need to be decided carefully, depending on the specific type of flow model employed. In particular, it should be noted that the direction of the wind approaching the wind farm may change depending on the total bottom resistance (and thus on $\beta$) due to, for example, the Coriolis effect. Such a change in the 'external' wind direction can be taken into account when the internal and external problems are coupled as in figure 3$a$, i.e., the correct external wind direction for a given $\beta$ can be calculated in the external flow model and returned to the internal model (as indicated by the dashed arrow in the left side of the figure). However, if the two sub-problems are decoupled as in figure 3$b$, we cannot correct the external wind direction in the internal problem for an updated $\beta$. The difference in the external wind direction between the cases with and without farm is, for most practical cases, expected to be relatively small. Nevertheless, such a change may still affect the array performance substantially and therefore need to be assessed carefully when the two sub-problems are decoupled as in figure 3$b$.

While the present theory describes a fundamental relationship between the turbine-scale and farm-scale flow problems and is therefore expected to serve as a basic framework for multiscale modelling of large wind farms in the future, there is still considerable room for improvement in the (especially analytical and low-order numerical) modelling of each sub-problem. In particular, it would be useful to develop an analytical model of $C_T^*$ that accounts for the effects of array layout and wind direction, especially for the case with a high array density, where the layout/direction effects are more complex due to combined effects of local blockage and wake mixing (Nishino and Draper 2019) and therefore the simple analytical model of $C_T^*$ given in (2.14) may yield a larger error. It would also be beneficial to further investigate the modelling of the momentum availability factor, $M$, under different types of atmospheric/weather conditions. Although the linear approximation model of $M$ proposed in (2.17) is useful for decoupling the external problem from the internal problem (and thus reducing the number of numerical simulations required for a given wind farm) as noted earlier in Section 2.4, the validity of such an approximation needs to be investigated further in future studies. All these improvements of flow models at each scale, combined following the present theory appropriately, would eventually enable more effective operation of existing large wind farms (using active control of turbine thrust and yaw angle, for example, for given weather conditions) and even a higher-level optimisation of future large wind farms, where the design of individual turbines may also be optimised for a given wind farm location (Nishino and Hunter 2018) to reduce their levelised cost of electricity (LCOE).

## 4. Conclusions

In this paper we have presented a fundamental theory based on the law of momentum conservation to help understand the complex multiscale problem of large wind farm aerodynamics. Care has been taken in the derivation of the theory to attempt to describe the basic relationship between the external (farm-scale) and internal (turbine-scale) flow problems in a generic manner so that the theory may be useful for various types and levels of large wind farm modelling, regardless of the specific details of flow models employed at each scale. In particular, unlike most of previous large wind farm models, the present theory does not consider modelling the ABL profile explicitly based on the logarithmic law. Instead, we have employed the concept of farm-average wind speed and derived a momentum equation that provides a generic coupling condition between the external and internal flow problems in terms of the reduction factor of the farm-average wind speed. This generic approach allows us to use the present theory in conjunction with a numerical weather model, for example, to investigate the effect of large-scale motions of the ABL on the aerodynamic performance of a large wind farm in a time-dependent (rather than statistical) manner. Although the theory has been rigorously derived only for a simplified wind farm situation where the flow over the turbine array is assumed to be in a fully developed state, we have also discussed how the theory can be applied (in an approximate manner) to real-world wind farm problems where the flow pattern around each turbine may vary over the entire farm. The theory therefore also provides a practical basis for estimating, e.g., the loss of power due to the so-called wind farm blockage effect.

To demonstrate how the present theory can help us determine the efficiency of a large wind farm from (modelled) solutions of the internal and external problems in a combined manner, we have also presented very simple analytical models for the local thrust coefficient $C_T^*$ (2.14) and the momentum availability factor $M$ (2.17), respectively. However, it should be remembered that the most important contribution of the present theory lies in the generality of the momentum equation (2.12) to be met in various types and complexity-levels of large wind farm modelling. Since the present theory provides only a framework for the coupling of the turbine-scale and farm-scale flow problems, the success of predictions of large wind farm performance still, of course, relies on the accuracy of flow models employed at each scale. While high-fidelity numerical flow models are already available at each scale, there is still considerable room for improvement in the analytical and low-order numerical modelling at each scale, which is necessary for high-level optimisation of future large wind farms. In particular, it would be helpful to further investigate and model in future studies how the local thrust coefficient $C_T^*$ is affected by various internal conditions (such as array configuration) in the internal problem, and how the momentum availability factor $M$ changes under various atmospheric/weather conditions in the external problem.

**Declaration of interests:** The authors report no conflict of interest.

## Appendix A. Effects of array configuration and wind direction

As noted earlier, one of the key issues in large wind farm modelling is how to consider the effects of array configuration and wind direction in a simple, inexpensive manner. To assess the validity of equation (2.14) introduced as the simplest possible model of $C_T^*$ (which does not account for either of these effects explicitly), we have carried out a series of LES of a realistic ABL flow over a periodic turbine array using the Met Office/NERC Cloud Model, called MONC (Brown et al. 2018; Hill et al. 2018). The domain is doubly periodic in the horizontal ($x$ and $y$) directions and has 128 grid points in the vertical ($z$) direction with a damping layer near the upper boundary. The dimensions of the domain are $\pi \times \pi \times 1$ (km) for $x$, $y$, and $z$, respectively. Individual turbines are modelled as actuator discs following the methodology of Calaf et al. (2010).

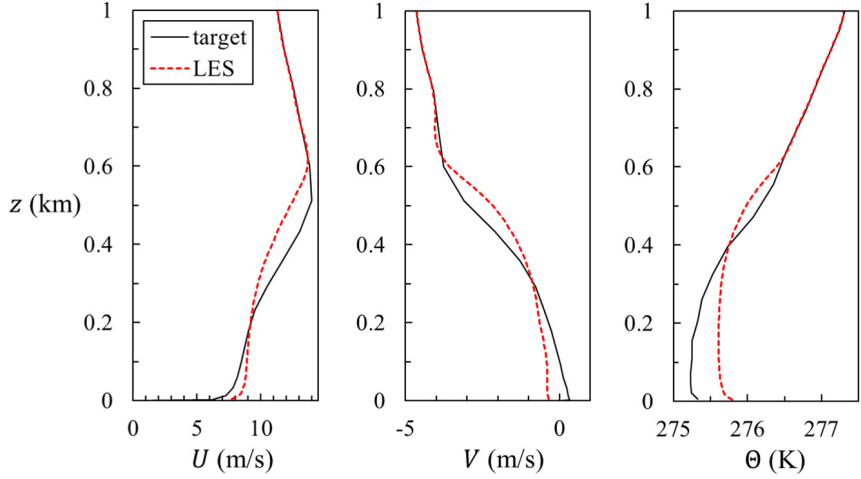

FIGURE 6. Target profiles extracted from UKV (solid lines) and mean profiles obtained from the reference LES without turbines (dashed lines) for $U$, $V$ and $\Theta$.

| case | configuration | $C_T'$ | $\Delta x/D, \Delta y/D$ |
|---|---|---|---|
| FA1 | fully aligned | 0.333 | 0.245 |
| FA2 | fully aligned | 0.667 | 0.245 |
| FA3 | fully aligned | 1 | 0.245 |
| FA4 | fully aligned | 1.333 | 0.245 |
| FS1 | fully staggered | 0.333 | 0.245 |
| FS2 | fully staggered | 0.667 | 0.245 |
| FS3 | fully staggered | 1 | 0.245 |
| FS4 | fully staggered | 1.333 | 0.245 |
| PS1 | partially staggered | 0.333 | 0.245 |
| PS2 | partially staggered | 0.667 | 0.245 |
| PS3 | partially staggered | 1 | 0.245 |
| PS4 | partially staggered | 1.333 | 0.245 |
| FS4(D) | fully staggered | 1.333 | 0.123 |

TABLE 1. Array configuration, turbine resistance and horizontal resolution for LES.

Simulations were run until a statistically stationary state was reached, and data were then collected over a period of approximately 4.5 hrs for analysis. To maintain statistical stationarity for a realistic ABL, a relaxation forcing term, $f_\varphi(z) = \left(\varphi_{\text{target}}(z) - \langle\varphi\rangle(z)\right)/\tau_{\text{relax}}$, was added to the governing equation for variable $\varphi$, where $\tau_{\text{relax}} = 3600s$ is a relaxation time scale and $\langle...\rangle$ represents an average over $x$ and $y$. This forcing was applied to horizontal velocities ($U$, $V$) and potential temperature ($\Theta$), with target profiles extracted from archived data of the Met Office's operational UK regional model (UKV: United Kingdom Variable) at a near-shore location in the North Sea. Figure 6 shows the target profiles for $U$, $V$ and $\Theta$ together with mean profiles obtained from LES without turbines. The target ABL profile was selected from a series of case studies conducted earlier (Dunstan et al. 2018) and represents a moderately unstable boundary layer typical of those observed at this near-shore location. The extracted wind profile was rotated so that $V = 0$ at hub-height ($H_{\text{hub}}$ = 100m) for the target profiles for $U$ and $V$; however, the stronger vertical mixing in the surface layer in the LES compared to the UKV means that this is not strictly maintained in the LES results, as can be seen in figure 6.

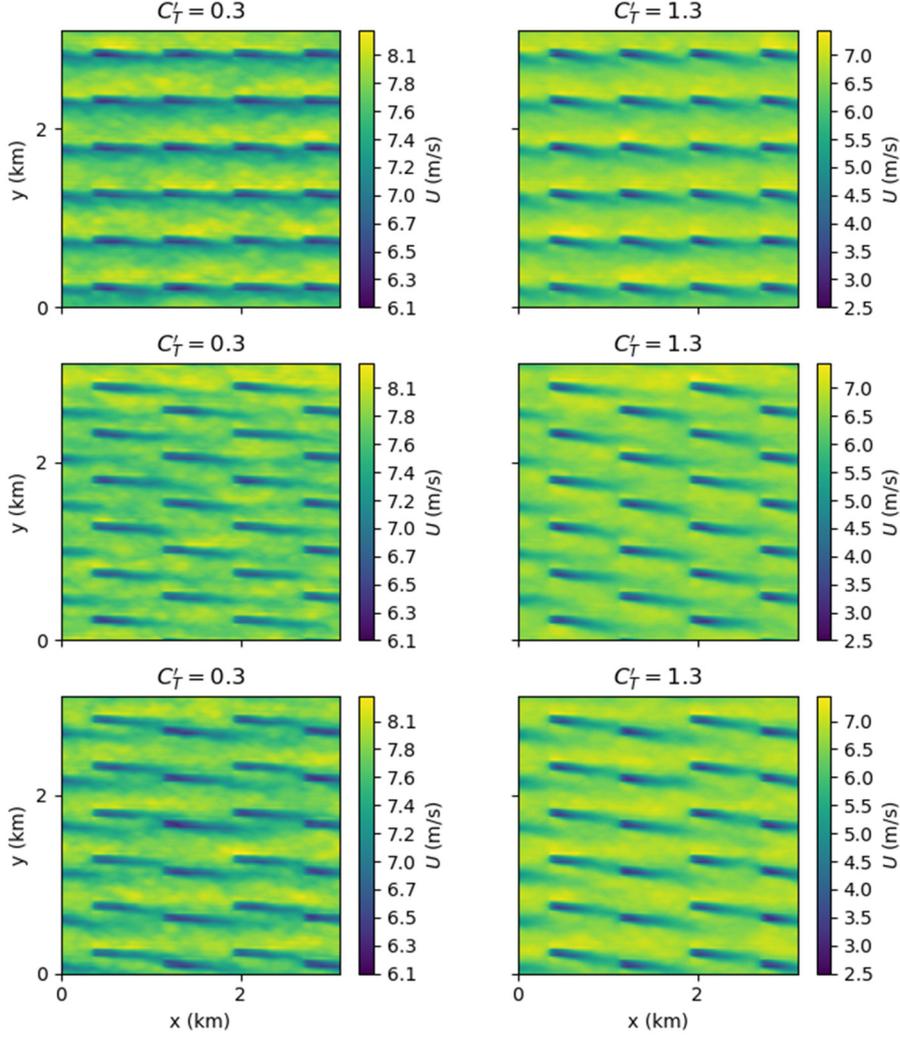

FIGURE 7. Time-averaged streamwise velocity contours at hub-height for fully aligned (top), fully staggered (middle) and partially staggered (bottom) arrays with $C_T' = 0.333$ (left) and 1.333 (right).

Note that the Coriolis force is not included in the present LES, although it is included in the UKV to produce the 'veered' target wind profile for the LES.

Three different turbine configurations were simulated: fully aligned, fully staggered, and partially staggered. A turbine diameter $D = 100$m was used for all simulations, with four different values for the turbine resistance coefficient, $C_T' = 0.333, 0.667, 1.0$ and $1.333$. The turbines were arranged in a $6 \times 4$ array with cross-stream spacing of $5.24D$ and streamwise spacing of $7.85D$. For the fully staggered and partially staggered cases, a cross-stream offset of $2.62D$ and $1.31D$, respectively, was applied to every other row. A summary of the configurations and settings used in the LES is given in Table 1. In addition to the 13 cases listed in Table 1, two reference cases with no turbines were run using both standard ($\Delta x = \Delta y = 0.245D$) and double ($\Delta x = \Delta y = 0.123D$) resolutions, which yielded two slightly different values ($2.44D$ and $2.38D$) for the farm-layer height $H_F$ from (2.8).

Examples of time-averaged $U$ contours on the horizontal plane at hub height are shown in figure 7 for all three array configurations. The deflection of turbine wakes (towards the negative $y$ direction, due to mixing down of veered wind from above) is evident in all cases, but especially so for higher values of $C_T'$ since the vertical mixing is more enhanced by turbines with higher resistance. For the highest $C_T'$ cases tested, the wakes are deflected by more than one turbine diameter before reaching the next row. Eventually, turbines in a fully aligned configuration, for example, are positioned in a high-speed flow region outside the wake of the turbine immediately upstream when $C_T'$ is high (as if an optimal 'wake steering' was performed to avoid direct wake interference).

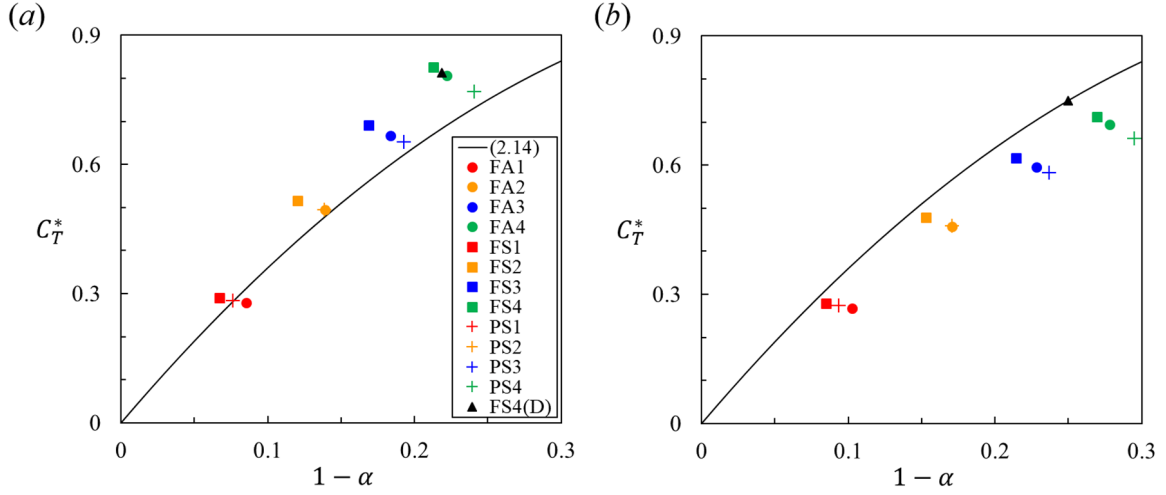

FIGURE 8. Relationship between $C_T^*$ $(= C_T' \alpha^2)$ and $1 - \alpha$ obtained from LES: (a) original results; and (b) corrected results following Shapiro et al. (2019); compared to theoretical prediction (2.14).

From the LES data, we have calculated the values of $\alpha = U_T/U_F$, where $U_T$ and $U_F$ are defined in (2.15) and (2.9a), respectively, and the local thrust coefficient $C_T^* = C_T' \alpha^2$ (note that $U_T$ is averaged over all 24 turbines in the domain). The results are plotted in figure 8(a) together with the simple theoretical prediction given earlier by (2.14). As can be seen, the values of $C_T^*$ obtained from the LES tend to be higher than the theoretical prediction, especially for high $C_T'$ cases. However, Shapiro et al. (2019) note that the values of $U_T$ derived from LES tend to be systematically overpredicted unless the resolution is very high, and propose a correction factor of the form:

$$m = \left(1 + \frac{C_T'}{4} \frac{\Delta}{\sqrt{3\pi} R}\right)^{-1}, \tag{A.1}$$

where $R = D/2$ is the disc radius and $\Delta$ is the effective filter size employed in LES. We have applied this correction factor to the value of $U_T$, using $\Delta = \sqrt{\Delta x^2 + \Delta y^2 + \Delta z^2}$, to produce corrected values of $\alpha$ and $C_T^*$ plotted in figure 8(b). The corrected $C_T^*$ values tend to be lower than the prediction given by (2.14), supporting the argument that this simple model may be used to estimate a practical upper limit to the value of $C_T^*$ (Nishino 2016). Comparing the three different array configurations tested, it can also be seen that the fully aligned configuration yields a lower $C_T^*$ value than the other two when $C_T'$ is low, but a higher $C_T^*$ value than the partially staggered configuration when $C_T'$ is high (since the wake deflection is significant enough to prevent direct wake interference as shown earlier), whereas the fully staggered configuration yields a higher $C_T^*$ value for the entire range of $C_T'$ tested. Overall, the difference in $C_T^*$ due to the (combined) effects of array configuration and wind direction is up to about 10% in these simulations. This level of change in $C_T^*$ is expected to be typical for real large wind farms where some but not all turbines may experience direct wake interference at a given time, although a larger discrepancy could be observed in some exceptional situations where all turbines in the farm are perfectly aligned with the wind direction; see, e.g., Porté-Agel et al. (2013).